# CONGESTION AVOIDANCE IN COMPUTER NETWORKS WITH A CONNECTIONLESS NETWORK LAYER PART I: CONCEPTS, GOALS AND METHODOLOGY

Raj Jain, K. K. Ramakrishnan
Digital Equipment Corporation
550 King Street (LKG 1-2/A19)
Littleton, MA 01460

## ABSTRACT

Congestion is said to occur in the network when the resource demands exceed the capacity and packets are lost due to too much queuing in the network. During congestion, the network throughput may drop to zero and the path delay may become very high. A congestion control scheme helps the network to recover from the congestion state.

A congestion avoidance scheme allows a network to operate in the region of low delay and high throughput. Such schemes prevent a network from entering the congested state. Congestion avoidance is a prevention mechanism while congestion control is a recovery mechanism.

We compare the concept of congestion avoidance with that of flow control and congestion control. A number of possible alternative for congestion avoidance have been identified. From these a few were selected for study. The criteria for selection and goals for these schemes have been described. In particular, we wanted the scheme to be globally efficient, fair, dynamic, convergent, robust, distributed, configuration independent, etc. These goals and the test cases used to verify whether a particular scheme has met the goals have been described.

We model the network and the user policies for congestion avoidance as a feedback control system. The key components of a generic congestion avoidance scheme are: congestion detection, congestion feedback, feedback selector, signal filter, decision function, and increase/decrease algorithms. These components have been explained.

The congestion avoidance research was done using a combination of analytical modeling and simulation techniques. The features of simulation model used have been described. This is the first report in a series on congestion avoidance schemes. Other reports in this series describe the application of these ideas leading to the development of specific congestion avoidance schemes.

Version: September 2, 1998

# 1 INTRODUCTION

Recent technological advances in computer networks have resulted in a significant increase in the bandwidth of computer network links. The ARPAnet was designed in the 1970s using leased telephone lines having a bandwidth of 50 Kbits/second. In the 1980s, local area networks (LAN) such as Ethernet and Token rings have been introduced with a bandwidth in the range of 10 Mbits/second. In this second half of the same decade, efforts are underway to standardize fiber optic LANs with a bandwidth of 100 Mbits/second and higher.

The steadily increasing bandwidth of computer networks would lead one to believe that network congestion is a problem of the past. In fact, most network designers have found the opposite to be true. Congestion control has been receiving increased attention lately due to an increasing speed mismatch caused by the variety of links that compose a computer network today. Congestion occurs mainly at routers (intermediate nodes, gateways, or IMPs) and links in the network where the rate of incoming traffic exceeds the bandwidth of the receiving node or link.

The problem of congestion control is more difficult to handle in networks with connectionless protocols than in those with connection-oriented protocols. In connection-oriented networks, resources in the network are reserved in advance during connection setup. Thus, one easy way to control congestion is to prevent new connections from starting up if congestion is sensed [1]. The disadvantage of this approach, like any other reservation scheme, is that reserved resources may not be used and may be left idle even when other users have been denied permission. Rather than get in to the *religious* debate between followers of the connection-oriented and connectionless disciplines, we simply want to point out the fact that the problem of congestion control in connectionless protocols is more complex. It is this set of protocols that we are concerned with here.

We are concerned with *congestion avoidance* rather than *congestion control*. The distinction between these two terms is a rather subtle one. Briefly, a congestion avoidance scheme allows a network to operate in the region of low delay and high throughput. These schemes prevent a network from entering the congested state in which the packets are lost. We will elaborate on this point in the next section where the terms flow control, congestion control, and congestion avoidance have been defined and their relationship to each other has been discussed.

We studied a number of alternative congestion avoidance shemes. This is the first report in a series about our work in this area. In this report, we discuss the goals, the metrics used to quantify the performance, and the fundamental components involved in the design of any congestion avoidance scheme. We address the issue of fairness in the service offered by a network. The role of algorithms for increase/decrease of the amount of traffic a user may place on the network is discussed, as well as the impact of the fairness of a range of increase/decrease algorithms. We also describe the simulation tools and the test sequences that we used to study alternative congestion avoidance schemes. These apply to all congestion avoidance schemes that we studied. Individual schemes will be described separately in other reports in this series.

The organization of this report is as follows. In section 2 we define the concepts of flow control, congestion control, and congestion avoidance. We discuss the distinction between these terms and show that the problem of congestion can not be solved simply by increasing memory size or increasing link bandwidth. Section 3 describes the requirements for an ideal congestion avoidance scheme. Section 4 lists a number of alternative schemes for congestion avoidance. From this list, we selected a few schemes for detailed study. The criterion for selection are described. Section 5 defines a number of performance metrics that were used to define optimality. Section 6 lists the goals that we set for the design of the schemes. Most of these goals are quantitative and verifiable. Section 7 describes the components of a generic congestion avoidance scheme in the framework of a control system. This helps in a systematic design of the scheme in that most components can be designed in isolation. Two component algorithms that are common to all schemes are window increase/decrease algorithms and window update frequency. These are described in sections 8 and 9, respectively. Finally, we describe the simulation model in section 10 and the test sequences used to verify the goals in section 11.

The next report in this series [19] describes a binary feedback congestion avoidance scheme.

# 2 CONCEPTS

In this section we define the basic concepts of flow control, congestion control, and congestion avoidance. These three concepts are related but distinct. They are related because all three solve the problem of resource management in the network. They are distinct because they solve resource problems either in different parts of the network or in a different manner. We also point out how decreasing cost of memory, or increasing link bandwidth and processor speed are not sufficient to solve these problems.

## 2.1 FLOW CONTROL

Consider the simple configuration shown in Figure 1a in which two nodes are directly connected via a link. Without any control, the source may send packets at a pace too fast for the destination. This may cause buffer overflow at the destination leading to packet losses, retransmissions, and degraded performance. A flow control scheme protects the destination from being flooded by the source.

Some of the schemes that have been described in the literature are window flow-control, Xon/Xoff [7], rate flow-control [5], etc. In the window flow-control scheme, the destination specifies a limit on the number of packets that the source may send without further permission from the destination. The permission may be an explicit message or it may be implicit in that an arriving acknowledgment may permit the source to send additional packets. The Xon/Xoff scheme is a special case of window flow-control in which the window is either infinity (Xon) or zero (Xoff). In the rate flow-control schemes, the destination specifies a maximum rate (packets per second or bits per second) at which the source may send information.

## 2.2 CONGESTION CONTROL

Now let us extend the configuration to include a network [1] (see Figure 1b) consisting of routers and links that have limited memory, bandwidth, and processing speeds. Now the source must not only obey the directives from the destination, but also from all the routers and links in the network. Without this additional control the source may send packets at a pace too fast for the network, leading to queuing, buffer overflow, packet losses, retransmissions, and performance degradation.

A congestion control scheme protects the network from being flooded by its users (source

---

[1]In most of the discussion on congestion, we use the term **network** to mean only the communication subnet not including the end nodes (called users).

and/or destination). [2]

In connection-oriented networks the congestion problem is generally solved by reserving the resources at all routers during connection setup. In connectionless networks it can be done by explicit messages (choke packets) from the network to the sources [16], or by implicit means such as timeout on a packet loss. Jain [14] and Ramakrishnan [18] have discussed a number of schemes for congestion control and analyzed a timeout-based scheme in detail.

## 2.3 FLOW CONTROL vs CONGESTION CONTROL

It is clear from the above discussion that the terms *flow control* and *congestion control* are distinct. Flow control is an agreement between a source and a destination to limit the flow of packets without taking into account the load on the network. The purpose of flow control is to ensure that a packet arriving at a destination will find a buffer there. Congestion control is primarily concerned with controlling the traffic to reduce overload on the network. Flow control solves the problem of the destination resources being the bottleneck while congestion control solves the problem of the routers and links being the bottleneck. Flow control is bipartite agreement. Congestion control is a social (network-wide) law. Different connections on a network can choose different flow control strategies, but nodes on the network should follow the same congestion control strategy, if it is to be useful. The two parties in flow control are generally interested in cooperating whereas the $n$ parties (e.g., different users) in congestion control may be noncooperative. Fairness is not an issue for the two cooperating parties whereas it is an important issue for $n$ competing parties.

It should be noted that there is considerable disagreement among researchers regarding the definitions of flow and congestion control. Some authors [7] consider congestion control to be a special case of flow control, while others [20] distinguish them as above.

## 2.4 CONGESTION AVOIDANCE

Traditional congestion control schemes help improve the performance after congestion has occurred. Figure 2 shows hypothetical graphs of response time and throughput of a network as the network load increases. If the load is small, throughput generally keeps up with the load. As the load increases, throughput increases. After the load reaches the network capacity, throughput stops increasing. If the load is increased any further, the queues start building, potentially resulting in packets being dropped. The throughput may suddenly drop when the load increases beyond this point and the network is said to be **congested**. The response time curve follows a similar pattern. At first the response time increases little with load. When the queues start building up, the response time increases linearly until finally, as the queues start overflowing, the response time increases drastically.

---
[2] We use the term **user** to denote source or destination transport entity. Either one could take action to limit the load on the network.

**Figure 2:** Network performance as a function of the load. Broken curves indicate performance with deterministic service and inter-arrival times.

The point at which throughput approaches zero is called the point of congestion collapse. This is also the point at which the response time approaches infinity. The purpose of a congestion control scheme [14, 3] is to detect the fact that the network has reached the point of congestion collapse resulting in packet losses and to reduce the load so the network can return to an uncongested state.

We call the point of congestion collapse a **cliff** due the fact that the throughput falls off rapidly after this point. If the goal of the network design is to maximize throughput while also minimizing response time, then the **knee** is a better operating point as shown in Figure 2. This is the point after which the increase in the throughput is small, but a significant increase in the response time results. Figure 2 also shows a plot of **power** [8] as a function of the load. Power is defined as the ratio of throughput to response time. The peak of the power curve occurs at the knee. We will discuss more about the use of power later under performance metrics.

A scheme that allows the network to operate at the knee is called a **congestion avoidance** scheme as distinguished from a *congestion control* scheme which tries to keep the network to the left of the cliff. A properly designed congestion avoidance scheme will ensure that the users are encouraged to increase their load as long as this does not significantly affect the response time and are encouraged to decrease it if that happens. Thus, the network oscillates around the knee and congestion never occurs. However, the congestion control schemes are still required to protect the network should it reach the cliff due to transient changes in the network.

## 2.5    CONGESTION AVOIDANCE vs CONGESTION CONTROL

The distinction between congestion control and congestion avoidance is similar to that between deadlock recovery and deadlock avoidance. Congestion control procedures are cures and the avoidance procedures are preventive in nature. A congestion control scheme tries to bring the network back to an *operating* state, while a congestion avoidance scheme tries to keep the network at an *optimal* state. Without congestion control a network may cease operating (zero throughput) whereas networks have been operating without congestion avoidance for a long time. The point at which a congestion control scheme is called upon depends on the amount of memory available in the routers, whereas, the point at which a congestion avoidance scheme is invoked, is independent of the memory size. A congestion avoidance scheme may continuously oscillate slightly around its goal (knee) without significant degradation in performance, whereas, a congestion control scheme tries to minimize the chances of going above the limit (cliff).

## 3    DESIGN REQUIREMENTS

Before we discuss the various schemes for congestion avoidance and compare them it is helpful to point out some of the design requirements that we followed. These requirements helped us limit the number of schemes for further study. The key requirements are: no control during normal operation, no extra packets, a connectionless network layer, and configuration independence. We describe these requirements below.

## 3.1 NO CONTROL DURING NORMAL OPERATION

Congestion is a transient phenomenon. Networks are configured in such a way that, on an average, the network is not overloaded. We therefore refrained from schemes that would generate extra overhead during normal (underloaded) conditions. This ruled out the use of such techniques as sending *encouragement packets* to users during underload and indicating overload by the absence of these packets.

## 3.2 NO NEW PACKETS

The processing overhead for network services depends upon the number of packets and the size of those packets. Performance measurements of existing implementations have shown that the number of packets affects the overhead much more than the size. Short acknowledgment messages cost as much as 50% of the long data messages. This is why piggybacking (combining two are more messages) helps reduce the overhead.

In summary, adding an extra packet causes much more overhead than adding a few bits in the header. We therefore preferred schemes that did not require generation of new messages and concentrated instead on adding only a few bits in the header.

## 3.3 DISTRIBUTED CONTROL

The scheme must be distributed and work without any central observer. Thus, schemes where all routers send congestion information to a central network control center were considered unacceptable.

## 3.4 CONNECTIONLESS NETWORK LAYER

The key architectural assumption about the networks is that they use connectionless network service and transport level connections. By this we mean that a router is not aware of the transport connections passing through it, and the transport entities are not aware of the path used by their packets. There is no prior reservation of resources at routers before an entity sets up a connection. The routers cannot compute the resource demands except by observing the traffic flowing through them.

Examples of network architectures with a connectionless network layer are DoD TCP/IP, Digital Network Architecture (DNA) [6], and ISO Connectionless Network Service used with ISO Transport Class 4 [9].

# 4 CONGESTION AVOIDANCE SCHEMES

Congestion control and congestion avoidance are dynamic system control issues. Like all other control schemes they consist of two parts: a feedback mechanism and a control mechanism. The feedback mechanism allows the system (network) to inform the users (source or destination) of the current state of the system. The control mechanism allows the users to adjust their load on the system. The feedback signal in a congestion avoidance scheme tells the users whether the network is operating below or above the knee. The feedback signal in a congestion control scheme tells the users whether the network is operating below or above the cliff.

The problem of congestion control has been discussed extensively in literature. A number of feedback mechanisms have been proposed. If we extend those mechanisms to signal operations around the knee rather than the cliff, we obtain a congestion avoidance scheme. Of course, the control mechanism will also have to be adjusted to help the network operate around the knee rather than the cliff. For the feedback mechanisms we have the following alternatives:

1. Congestion feedback via packets sent from routers to sources.
2. Feedback included in the routing messages exchanged among routers.
3. End-to-end probe packets sent by sources.
4. Each packet contains a congestion feedback field that is filled in by routers in packets going in the reverse direction.
5. A congestion feedback field is filled in by routers in packets going in the forward direction.

The first alternative is popularly known as *choke packet* [16] or *source quench message* in ARPAnet [17]. It requires introducing additional traffic in the network during congestion, which may not be desirable. A complement to this scheme is that of encouraging sources to increase the load during underload. The absence of these *encouragement messages* signals overload. This scheme does not introduce additional traffic during congestion. Nevertheless, it does introduce control overhead on the network even if there is no problem.

The second alternative, increasing the cost (used in the forwarding database update algorithm) of congested paths, has been tried before in ARPAnet's delay-sensitive routing. The delays were found to vary too quickly, resulting in a large number of routing messages and stability problems. Again, the overhead was not considered justifiable [15].

The third alternative, probe packets, also suffers from the disadvantage of added overhead unless probe packets had a dual role of carrying other information in them. If the latter were the case, there would be no reason not to use every packet going through the network as a probe packet. We may achieve this by reserving a field in the packet that is used by the network to signal congestion. This leads us to the last two alternatives.

The fourth alternative, reverse feedback, requires routers to piggyback the signal on the packets going in the direction opposite the congestion. This alternative has the advantage in that the feedback reaches the source faster. However, the forward and reverse traffic are not always related. The destinations of the reverse traffic may not be the cause of or even the participant in the congestion on the forward path. Also, many networks (including DNA) have path splitting such that the path from A to B is not necessarily the same as that from B to A.

The fifth alternative, forward feedback, sends the signal in the packets going in the forward direction (direction of congestion). The destination either asks the source to adjust the load or returns the signal back to the source in the packets (or acknowledgments) going in the reverse direction. This is the alternative that we finally chose for further study.

The minimal forward feedback requires just one bit of feedback signal with every packet. Although at first, one bit may not appear to be able to carry enough information, we show in the second part [19] of this report series that there is considerable performance gain even by single-bit feedback.

Most of the discussions in this and associated reports center around window-based flow-control mechanisms. However, we must point out that this is not a requirement. The congestion avoidance algorithms and concepts can be easily modified for other forms of flow control such as rate-based flow control in which the sources must send below a rate (packets/second or bytes/second) specified by the destination. In this case, the users would adjust rates based on the signals received from the network.

# 5 PERFORMANCE METRICS

The performance of a network can be measured by several metrics. The commonly used metrics are: throughput, delay, and power.

Throughput is measured by the user bits transmitted per unit of time. Thus, protocol overhead, retransmissions, and duplicate packets are not considered in throughput computation.

Some of the more important applications of computer networks are: file transfer, mail, and remote login. The first two are throughput sensitive. The response time (time for the packet to reach the destination) is generally not so important. On the other hand, for remote login, response time is more important than throughput.

The aforementioned goal, maximizing throughput and minimizing response time, are mutually contradictory in that all methods to increase throughput result in increased response time as well and vice versa. To resolve this contradiction, Giessler et al. [8] proposed the following metric:

$$\text{Power} = \frac{\text{Throughput}^\alpha}{\text{Response time}}$$

Here, $\alpha$ is a positive real number. Notice that by maximizing power, one tries to maximize throughput and minimize response time. Normally, $\alpha = 1$, i.e., increasing throughput and decreasing response time are given equal weights. By setting $\alpha > 1$, one can favor file traffic by emphasizing higher throughput. Similarly, by setting $\alpha < 1$ one can favor terminal traffic by emphasizing lower response time.

It must be pointed out that the throughput and response time used above are system-wide throughput (total number of packets sent for all users divided by the total time) and system-wide response time (averaged over all users) giving us *system power*. The operating point obtained in this manner is different from the one that would be obtained if each of the $n$ users tries to maximize their own *individual power* (ratio of individual throughput and individual response time). Maximizing individual power leads to a number of undesirable effects [2, 10].

# 6  GOALS

Design of a congestion avoidance scheme requires comparing a number of alternative algorithms and selecting the right parameter values. To help us do this we set a number of goals which are described in this section. Each of these goals has an associated test to help us verify whether a particular scheme meets the goal.

## 6.1  EFFICIENT

A network operating at the knee is said to be operating efficiently. The efficiency is measured by the *system power* as defined earlier. The congestion avoidance scheme should lead the network to the knee, that is, the point of maximum system power.

Given any performance metric and a system of $n$ users, there are two kinds of efficient operating points: individual and global. Individually efficient operating points occur when each user tries to maximize its performance without regard for the performance of others. This may or may not lead to the globally efficient operating point where the total system performance is the highest. In other words, at the globally efficient operating point, there may still be opportunities for each individual user to improve its performance (while degrading that of others). We have explicitly chosen *global efficiency* and not individual efficiency as our goal.

We set the parameters of our congestion avoidance schemes to values that maximize global power and fairness.

## 6.2 RESPONSIVENESS

Network configurations and traffic vary continuously. Nodes and links come up and down. The load placed on the network by users is highly varying. The optimal operating point is therefore a continuously moving target. It cannot be assumed that the optimal operating point observed in the past is still optimal because the configuration or workload might have changed. If the feedback is limited to a binary signal [19], this leads to schemes that continuously change the load slightly below and slightly above the optimal level and verify the current state by observing feedback signals obtained from the network.

When operating at the knee, to sense the state of the network, there is a need for oscillation of the window around the optimal level. Any attempt to eliminate oscillations also leads to the loss of responsiveness. We explicitly tested the responsiveness of the algorithms by changing router service times during a simulation and verifying that the operating point followed the change in the optimal window size.

## 6.3 MINIMUM OSCILLATION

Schemes with a smaller amplitude of oscillation are preferable over those with a larger amplitude. We found that schemes with smaller oscillations are also slower (less responsive) algorithms in that they take longer to reach the target. We therefore need to make a suitable tradeoff between the two requirements.

## 6.4 CONVERGENCE

If the network configuration and workload were to remain stable, the scheme should bring the network to a stable operating point. Many alternatives were rejected because they were divergent. This means that the total load on the network either increased slowly towards infinity or decreased towards zero without stabilizing.

We also found cases where the system throughput would converge to a stable value but the individual user's throughput would vary considerably. We call this phenomenon *false convergence*. More specifically, it could be called global convergence without local convergence.

Among converging schemes, the preferred alternative is the one that takes the least time to converge.

## 6.5 FAIRNESS

In any system shared by $n$ independent users, fairness is an important issue. Fairness demands that given $n$ users sharing a resource, each of them should get the same share of the resources (unless the user itself demanded less than its fair share). Thus, $n$ users sharing a path and each demanding infinite resources should have equal throughput. If the throughputs are not exactly equal, the fairness can be quantified using the following fairness function [11]:

$$\text{Fairness} = \frac{(\sum_{i=1}^{n} x_i)^2}{n \sum_{i=1}^{n} x_i^2}$$

Here, $x_i$ is $i^{th}$ user's throughput.

In designing congestion avoidance schemes and in setting parameter values our goal was to choose schemes and values that maximized fairness. Often, we found that there is a tradeoff between efficiency and fairness. The values that maximize system power are not necessarily the same as those that maximize fairness and vice versa. In such cases, we tried to err on the side of efficiency.

The definitions of fairness and efficiency presented in this report treat the network as a single resource to be shared equally among all users.

## 6.6 ROBUSTNESS

Robustness requires that the scheme work in a noisy (random) environment. Thus, schemes which work only for deterministic service times or schemes that presume a particular dis-tribution (exponential) for service times were discarded. We tested robustness by varying distributions of service times.

## 6.7 SIMPLICITY

Simplicity of schemes is also an important goal. For most alternatives we tried their simpler versions. Only if the simpler versions caused a significant reduction in performance did we sacrifice simplicity.

## 6.8 LOW PARAMETER SENSITIVITY

In designing the congestion avoidance schemes we studied sensitivity with respect to parameter values. If the performance of a scheme was found to be very sensitive to the setting of a parameter value, the scheme was discarded.

## 6.9 INFORMATION ENTROPY

Information entropy relates to the use of feedback information. We want to get the maximum information across with the minimum amount of feedback. Given $n$ bits of feedback, information theory tells us that the maximum information would be communicated if each of the $2^n$ possible combinations were equally likely. In particular, with one bit of feedback, maximum information would be communicated if the bit was set 50% of the time, i.e.,

$$P(bit = 1) = P(bit = 0) = 0.5$$

## 6.10 DIMENSIONLESS PARAMETERS

A parameter that has dimensions (length, mass, time) is generally a function of network speed or configuration. A dimensionless parameter has wider applicability. For example, in choosing the increase algorithm we preferred to increase the window by an absolute amount of $k$ packets rather than a rate of $t$ packets/second. The optimal value of the latter depends upon the link bandwidth. Our goal in developing the congestion avoidance scheme was to have all parameters dimensionless, thereby ensuring that the scheme would be applicable to networks with widely varying bandwidths. Of course, we also studied the parameter sensitivity and chose the least sensitive alternatives.

**Figure 3:** Components of a congestion avoidance scheme

## 6.11 CONFIGURATION INDEPENDENCE

Configuration independence is a desirable goal. We therefore tested our schemes for many different configurations. Although it is a noble goal, generally it is possible to come up with a configuration where a given scheme will not satisfy one of the goals. We have tried to identify such limitations wherever appropriate.

In many network architectures, including DNA, paths are dynamically calculated. As the routers go up or down, the paths change. The congestion avoidance scheme should adapt to the changing configurations.

# 7  COMPONENTS OF AN AVOIDANCE SCHEME

The two key components of any congestion avoidance scheme, the feedback mechanism and the control mechanism, have already been discussed earlier in this report. We call these **network policies** and **user policies**, respectively. A more detailed break down of these policies is shown in Figure 3. This allows us to concentrate on one component at a time and test various alternatives for that particular component. During the analysis, it can be assumed that other components are operating optimally. Of course, one would need to verify at the end that the combined system worked satisfactorily under imperfect conditions.

The network policy consists of three algorithms: congestion detection, feedback filter, and feedback selector. The user policy also consists of three algorithms: signal filter, decision function, and increase/decrease algorithm. These generic algorithms apply to many different congestion avoidance schemes. For example, these six algorithms would apply whether we choose to implement network feedback in the form of source quench messages or we implement it via a field in the packet header.

## 7.1 CONGESTION DETECTION

Before the network can feedback any information, it must determine its state or load level. In a general case, the network may be in one of $n$ possible states. The congestion detection function helps map these states into one of the two possible load levels: overload or underload (above or below the knee). A k-ary version of this function would result in $k$ levels of load indications. A congestion detection function, for example, could work based on the processor utilization, link utilization, or queue lengths.

## 7.2 FEEDBACK FILTER

After the network has determined the load level, it may want to verify that the state lasts for a sufficiently long period before signaling it to the users. This is because a feedback of state is useful only if the state lasts long enough for the the users to take action based on it. A state which changes very fast may lead to confusion. By the time users become aware of the state, it no longer holds and the feedback is misleading. Therefore, we need a (low-pass) filter function to pass only those states that are expected to last long enough for the user action to be meaningful. Examples of feedback filters are exponential weighted average or moving average of processor utilization, link utilization, or queue lengths.

## 7.3 FEEDBACK SELECTOR

After the network has determined that it is overloaded (or underloaded) and has ensured that the state is likely to last long enough, it needs to communicate this information to users so that they may reduce (or increase) the traffic. A feedback selector function may be used to determine the set of users to be notified. In other words, the network may want all users to reduce the traffic or it may selectively ask some users to reduce and others to increase the traffic. In the simplest case, it may give the same feedback signal to all users.

## 7.4 SIGNAL FILTER

The users receiving the feedback signals from the network (routers) need to interpret the signal. The first step in this process is to accumulate a number of signals. Due to the probabilistic nature of the network, all these signals may not be identical. Some may indicate that the network is overloaded while others may indicate that it is underloaded. The user needs to combine these to decide its action. Some examples of received signal filter are *majority voting* (50%), or *three-quarter majority* (75%), or *unanimous* (100%). The percentage may be used after applying a weighting function, for example, giving higher weight to recent signals.

## 7.5 DECISION FUNCTION

Once the user knows the network load level, it has to decide either to increase its load or decrease its load. The function can be broken down into two parts: the first part determines the direction and the other determines the amount. These parts are called decision function and increase/decrease algorithms, respectively.

The decision function takes feedback signals for the last $T$ seconds, for instance, as input parameter, and determines the load level of the network path. The key parameter is $T$ - , the interval for which it should accumulate feedback. This determines the window update frequency. We will further discuss window update frequency later in this report.

In its simplest form a decision function may be a 2-way function indicating whether the load should be increased or decreased. Some would argue that it may be a 3-way function including a *gray area* where no action is taken.

Another generalization often mentioned is to make a decision but not act on it unless we reach the same decision again, one or more times in the future. This may seem to increase the probability of reaching the right decision.

Both the generalizations mentioned above result in *postponement* of the action thereby causing the system to stay in the same state longer. This may be useful if the goal (knee) is stable but in a computer network the knee is a continuously moving target and it is helpful to reconfirm the state by perturbing the load, however slightly, one way or the other.

The costs of the two types of errors, increasing the traffic under overload and decreasing the traffic under underload, determine whether the users should err on the side of being pessimistic or optimistic. For cliff-based policies (congestion control schemes), it is better to be pessimistic because the cost (loss of a packet) of miss-signal is high. For knee-based policies (congestion avoidance schemes), the two costs are approximately equal (assuming the knee is far away from the cliff). We therefore recommend a two-way decision function and no postponement of action.

## 7.6   INCREASE/DECREASE ALGORITHM

The key part of a control scheme is the control, i.e., the action taken as a result of the feedback. For congestion avoidance schemes this part lies in the increase/decrease algorithms used by the users. These algorithms are a key to achieving efficiency as well as fairness. The choice of other components of the congestion avoidance scheme depends upon the type of feedback chosen, whereas, the increase/decrease algorithms can be discussed and analyzed generically in great detail and apply to several feedback mechanisms. We discuss some of these alternatives in the next section. A more complete discussion may be found in Chiu and Jain [4].

# 8   INCREASE/DECREASE ALGORITHMS

In this section we compare a number of alternative algorithms for window increase and decrease. We show that an additive increase, multiplicative decrease algorithm provides fair and stable operation and that it is important to keep windows as real valued variables which are rounded-off to the nearest integer.

We assume that the source and destination transport entities are using a window-based flow-control. Thus, increasing the window increases the load on the network and decreasing the window decreases the load. It must be pointed out, however, that all the arguments apply equally well to other forms of flow control such as rate based flow-control, in which the destination permits the source to send data at a pre-specified rate (bits/second or packets/second). In this case, it is obvious that increasing the rate increases the load and vice versa.

A general increase (or decrease) algorithm would take the current control (flow-control window) and feedback signals as input arguments and produce the new control as an output argument. However, as discussed above, we assume that the feedback signals have been analyzed by other components of the congestion avoidance scheme and the decision provided to this component is to increase or decrease the traffic. Thus, the key parameter to the increase/decrease algorithms is the current window.

We considered two types of increase/decrease algorithms:

1. Additive - The window is increased or decreased by a fixed amount.

$$w = w + k_1$$

$$w = w - k_2$$

2. Multiplicative - The window is increased or decreased by a fixed multiple.

$$w = r_1 w, r_1 > 1$$

$$w = r_2 w, 0 < r_2 < 1$$

More general increase/decrease algorithms using linear and non-linear functions of the window were also considered and are described in Part 3 of this report series [4]. Here, we concentrate on choosing one of the following four combinations:

1. Multiplicative Increase, Multiplicative Decrease
2. Multiplicative Increase, Additive Decrease
3. Additive Increase, Additive Decrease
4. Additive Increase, Multiplicative Decrease

In all these alternatives we assume that the computed value is *rounded* to an integer value and that the window is never allowed to go below 1.

The two key requirements of the increase/decrease policy are that it should allow a single user network to operate as close to optimality as possible and that it should allow a multi-user network to operate as fairly as possible. In comparing the above alternative we will assume a simplified model of the network in which all users share the same path and therefore receive the same feedback. If $i^{th}$ user has a window $w_i$, the network gives the signal to go up if and only if:

$$\sum_{i=1}^{n} w_i \leq w_{knee}$$

Here, $w_{knee}$ is the window at the knee of the throughput (or response time) curve for the given network configuration.

The fairness goal dictates that regardless of any starting point all $n$ users should converge to the same final window $w_{knee}/n$. While going down, the users with higher windows should go down more than those with lower windows, i.e., the decrease should be proportional (multiplicative). While going up, the users with lower windows should go up more than

**Figure 4:** Different increase/decrease algorithms may lead to fair or unfair stable operating points.

those with higher windows, i.e., the increase cannot be multiplicative. These observations leave us only with the fourth alternative of additive increase and multiplicative decrease. The other three alternatives are unfair, that is, they may stabilize at points where the windows are not equal. Instead of proving it mathematically, we show an example in Figure 4. We consider a network shared by two users. The optimal window for this network configuration $w_{knee}$ is assumed to be 15.5. User 1 starts first and User 2 joins in later. Figure 4b shows the window sizes for the two users (called **window trajectories**) with both users following an *additive increase by 1 and additive decrease by 1* algorithm. After a while the two users stabilize with the following sequence of window sizes and feedback signals:

|  |  |  |  |  |
|---|---|---|---|---|
| User 1 Window: | 14 | 15 | 14 | ... |
| User 2 Window: | 1 | 2 | 1 | ... |
| Total Window: | 15 | 17 | 15 | ... |
| Network Signal: | up | down | up | ... |

Thus, the system reaches a *stable* state where the first user oscillates with an average window of 14.5, while the second user oscillates with an average window of 1.5. The throughput of the first user is approximately ten times that of the second. The algorithm is unfair.

It can similarly be shown that the first two alternatives with multiplicative increase are unfair.

The window trajectory, using the fourth alternative (additive increase and multiplicative decrease) for the same network configuration, is shown in Figure 4c. With this algorithm the two users stablize at window size very close to each other. This algorithm is fair in most cases. The unfairness occurs in this case mainly due to the control (window) being discrete (integer valued). This issue as well as a few others are discussed next.

## 8.1 EFFECT OF DISCRETE CONTROL

An important aspect of increase/decrease algorithms is the truncation/rounding issue. There are two values for the window size that a user maintains: computed and implemented. The values computed using the increase and decrease algorithms are real valued variables. However, if the computed window comes out to 2.6, for instance, the user must decide whether to use 2 (truncated) or 3 (rounded) as the number of packets (implemented window) which will be sent in the next cycle.

It is important that the window values be maintained as real numbers even though actual windows used are integer valued. It is possible to study the effect of discrete control by studying the variation of the window at a user in isolation assuming that the network feedback is perfect (based on the global knowledge). Thus, using a simple computer program, it is possible to try various values of increase amount, decrease factor, starting window values, and to find the cases where the algorithms stabilize to unfair values. We found that generally, single precision floating point representation of window is adequate.

If only integer values are maintained for the window, *additive increase and multiplicative decrease* may also stabilize to unfair values, although this may not be the case for all values of increase amounts and decrease factors. Figure 5 shows a particular case of unfairness from having discrete control. The two-user configuration discussed earlier in Figure 4 is used. The users increase additively by 1 and decrease multiplicatively by a factor of 0.8. The optimal window is 15.5. After a while the two users stabilize such that User 1 has a window of 10 and user 2 has a window of 6. The sum is more than $w_{knee} = 15.5$ and therefore both users are asked to reduce. They come down (using a factor of 0.8) to 8 and 4 (0.8(6)=4.8 truncated to 4). The total window is less than $w_{knee}$ and hence both users are asked to go up. They go up by 1 to 9 and 5. The total window is still less than $w_{knee}$ and the users go up to 10 and 6. After this, the cycle repeats and the second user gets 6/10th of the first user's throughput.

**Figure 5:** An example of unfairness caused by discrete (integer valued) window sizes even with additive increase and multiplicative decrease.

The same configuration is fair when rounding is used, as was illustrated earlier in Figure 4c. By exhaustively searching the parameter space, we verified the fairness of the additive increase and multiplicative decrease algorithm when the implemented window size is obtained by rounding the computed window. A rigorous mathematical analysis of various alternatives is presented in [4].

## 8.2   INCREASE AMOUNT AND DECREASE FACTOR

The additive increase and multiplicative decrease with rounding lead to a fairly stable operation for all values of the two parameters, namely, the increase amount and the decrease factor. However, not all values are equally good. The values affect the time required to converge to a stable operation and the amount that the total traffic on the network will oscillate during stable operation. The goal is to minimize the time to convergence as well as to minimize the oscillation size during stable operation. Unfortunately, these two goals are contradictory in the sense that the parameter values that decrease the time needed to reach stable operation tend to increase the oscillation size also.

We recommend using an increase amount of 1 and a decrease factor of 0.875. The first value was chosen to minimize the size of oscillations and also to ease computations on a wide variety of processors. Multiplying by 0.875 ($1 - \frac{1}{8}$) requires an arithmetic shift operation and subtraction.

**Figure 6:** Birth policies.

## 8.3 BIRTH POLICIES

Another alternative to convergence time and the oscillation size dilemma is the use of a birth policy. The parameter values are initially chosen to minimize the time to convergence. Once convergence is reached, another set of parameter values is used which minimizes the oscillation size. The convergence is detected by a change of direction (a decrease following an increase or vice versa).

Figure 6 shows the case of a single user passing through a path with a knee at 15.5, starting with a window of 1. Two cases are shown. In the first case (without a birth policy) the increase amount and the decrease factor are set at 1 and 0.875, respectively. In the second case (with a birth policy) the increase amount is 2 until the first decrease when the increase amount is reset to 1. It is seen that the birth policy does allow the user to reach the knee faster. However, the additional complication of keeping an additional code to detect the direction change may not be considered worthwhile.

## 8.4 SOURCE BOUND CASE

In the discussion so far, we have assumed that the sources are able to send as many packets as the optimal window computation requires. An interesting case to consider is what happens if the source (and not the network or the destination) is the bottleneck. In this case, the network always gives increase signals to the user which computes a new larger window but is not able to send more than $w$ packets in one round-trip delay. Based on $w$ packets per cycle, the network continues to ask the user to increase the load. In this case, it is possible for the computed window to increase continuously and overflow. Actually, in this case the computed window has no meaning and therefore should never be increased beyond $w + dw$, where $dw$ is the increase amount and $w$ is the previously 'used' window. This leads to the rule that a user does not increase the window if it has not been able to implement the previous increase.

## 8.5 DESTINATION BOUND CASE

If the destination has limited buffering then it can impose a limit on the window used by the source. The source should never increase the window beyond that permitted by the destination. It tries to satisfy both the destination as well as the network.

## 8.6 K-ARY SEARCH

In addition to the four alternatives of additive/multiplicative, increase/decrease, we also tried a k-ary search for the operating point determined by the knee. The well-known binary search is a special case of k-ary search with k=2. In the k-ary search, the user remembers the highest and lowest windows at which the direction was changed. A direction change is defined as an increase followed by a decrease or vice versa. If $w_{low}$ and $w_{high}$ are the two window values at which the direction was changed, the user next tries the window:

$$w = w_{low} + \frac{w_{high} - w_{low}}{k}$$

Here, $k$ is a real number greater than 1.

We found that the k-ary search not only requires additional state variables ($w_{low}$ and $w_{high}$) to be maintained, but it also is less responsive. It works fine in stable configurations. However, in cases where the number of users or router speeds change during simulation, we need algorithms to allow previously remembered values to be forgotten in favor of the new information. This introduces additional complexity making the k-ary search not worthwhile pursuing.

The discussion on increase/decrease is summarized by the algorithm given in Box 1.

# 9 WINDOW UPDATE FREQUENCY

The issue of window update frequency involves a decision on how often the users should change their windows. This is another component algorithm (along with increase/decrease)

Box 1: Increase/decrease algorithms

```
PROCEDURE increase(w,w_max,w_used);
REAL w;                                         !Computed window (real valued);
INTEGER w_max;                                  !Window permitted by destination;
INTEGER w_used;                                 !Window used (integer valued);
BEGIN
        w:=w + 1;                               !Go up by 1;
        IF (w > (w_used + 1)) THEN w := w_used + 1; !No more than 1 above last used;
        IF (w > w_max) THEN w := w_max;         !Obey destination;
        w_used:=Entier(w + 0.5);                !Entier(x) gives an integer ≤ x;
END of increase;

PROCEDURE decrease(w,w_used);
REAL w;                                         !Computed window (real valued);
INTEGER w_used;                                 !Window to be used (integer valued);
BEGIN
        w := 0.875 * w;                         !Multiplicative decrease;
        IF (w < 1) THEN w := 1;                 !Do not reduce below 1;
        w_used:=Entier(w + 0.5);                !Round-off;
END of decrease;
```

that is common to many different congestion avoidance schemes that we considered. The ideas that are common to all avoidance schemes are being discussed here. A more specific discussion relating to the binary feedback scheme appears in [19].

The key results we want to present in this section are that windows should be adjusted once every two round-trip delays (two window turns) and that only the feedback signals received in the past cycle should be used in window adjustment. We present a set of control theoretic arguments and show that the simplest control scheme is obtained with these two restrictions.

In every system control scheme, we have a choice of exercising control (e.g., changing a window) every time we have a new feedback signal from the system. Exercising control too often may lead to unnecessary oscillations, while delaying control for long may lead to a lethargic system that takes too long to converge. The optimal control frequency depends upon the feedback delay, that is, the time required for the control to take effect.

To demonstrate the feedback delay, consider an example of a new source of traffic deciding to join the network with a large starting window of $w_1$. As shown in Figure 7, this is a case of the source changing its window from $w_0$ to $w_1$ with $w_0 = 0$. Let us assume that this happens at time $t = 0$. The effect of this window change will not be felt immediately. In fact, the first few packets will find the network response to be the same as before the source came on. The first network feedback to the source will come with the first packet at time $t = r_0$, where $r_0$ is the round-trip delay corresponding to the old control (zero window from this source). It is only the first packet in the next window cycle $((w_1 + 1)^{th}$ packet) that will bring a network feedback corresponding to window $w_1$. This packet would enter the network at time $t = r_0$ and come back at time $t = r_0 + r_1$, where $r_1$ is the round-trip delay corresponding to window $w_1$. The key point to notice is that it takes at least [3] two round-trip delays for the effect of a window change to be observed. The feedback signals $\mathbf{y}(n)$ (a vector) observed in the $n^{th}$ cycle correspond to the windows during cycles $n - 1$ and $n - 2$.

**Figure 7:** Decision Frequency. After the window $w$ is changed from $w_0$ to $w_1$, the feedback **f** received during the second round-trip delay interval is a function of $w_1$. That received during the first round-trip delay is a function of both $w_0$ and $w_1$.

$$\mathbf{y}(n) = fn\{w(n-1), w(n-2)\}$$

Here, $w(n)$ is the window in cycle $n$. It may be determined as a function of all past feedback and window history:

$$w(n+1) = fn\{w(n-j), \mathbf{y}(n-i), i = 0, 1, 2, \ldots, j = 0, 1, 2, \ldots\}$$

Once we understand the delayed feedback aspect, it is possible to write down the state space equations for the system and determine the optimal control policy. The most general control functions may require us to remember a long history. The simplest control policy is obtained if we keep the window constant for two cycles, so that $w(n-1) = w(n)$ for $n$ even, and use only the feedback for the last cycle, that is, for even values of $n$:

$$\mathbf{y}(n) = fn\{w(n-1)\}$$

$$w(n+1) = fn\{w(n), \mathbf{y}(n)\}$$

---

[3] The delay may be more if the network feedback signals are based on the state of the network in the previous cycle rather than this cycle.

This is the argument for adjusting the window only every two round-trip delays and for using the feedback signals obtained during the last round-trip interval.

# 10 SIMULATION MODEL

Various schemes for congestion avoidance were studied using a combination of analytical and simulation modeling techniques. In this section, we describe the simulation model used. The model described in [12], has been extended to simulate congestion avoidance algorithms. The model written in SIMULA simulates portions of network and transport layers and was used initially to study timeout algorithms [13] and a timeout based congestion control scheme [14]. In the model, the transport layer protocol is simulated in detail. In particular, the flow control portion is a verbatim copy of the DNA's transport layer specifications [6]. The routers and links are modeled as a single queue.

A number of features of the model including the trace facility and various modules have already been described in [12], and will not be repeated here. Instead we concentrate on the new features: shared paths, non-homogeneous paths, satellite links, and staggered start. Brief descriptions of these features follow.

## 10.1 SHARED PATHS

The model allows a configuration shown in Figure 8a. A number of sources on one local area network communicate with a number of destinations on another local area network and connect to the first by a number of routers. Its logical representation is shown in Figure 8b. There is no limitation on the number of users or number of routers.

## 10.2 NON-HOMOGENEITY

Routers and links are represented by queues, each of which can have a different service rate. The service time is determined by the packet length and the service rate. Local traffic is assumed to slow down the service rate. Thus, even if all routers and links are operating at the same speed, different service rates may be used depending upon the percentage of resources used up by local traffic.

**Figure 8:** Initial configurations. Two LANs interconnected via lower speed links possibly including a satellite link.

## 10.3 SATELLITES

Satellite links consist of a ground station followed by a long propagation delay. The ground station is simulated like other intermediate nodes, that is, a queue with a given service rate. The propagation delay is constant. A path may contain any number of satellite links.

## 10.4 STARTING GAP

It is possible to specify starting times individually for each user. This helps us verify that the scheme adapts to a changing workload as users go on/off the network.

## 10.5 TRANSIENTS

It is possible to simulate a transient change in service rate caused by increased local traffic or by nodes/links going down and coming back up. The specified simulation length ( of packets to send) is divided into three parts. During the middle part, the service rate of the bottleneck, as well as number of buffers there, may be changed (increased/decreased) to any specified value.

## 10.6 PARAMETER VARIATION

There are more than 40 different input parameters to the model. It is possible to run many simulation experiments by specifying a range of values of any subset of the parameters. This allows us to get a plot of any performance variable as a function of the parameter values, thereby finding the optimal value as well as determining the sensitivity.

## 10.7 WORKLOAD

The workload is specified by the number of active users and the packet size distribution. Once a user is active, it has a sufficient supply of packets to transmit, and is limited only by the flow-control windows. The packet sizes can take a number of different distributions such as: constant, uniform, exponential, bimodal, erlang, etc. Constant values are useful in running deterministic simulations and debugging.

By specifying the network service time that is less than the source service time, the model allows the simulation of a case where the throughput is limited by the source and not the network. However, this case is not very interesting, as the network has little congestion. We use this case to verify that all the algorithms do work in the source-bound case also.

## 10.8 PERFORMANCE DATA

The model provides time-varying graphs of the performance of individual users, routers, as well as the whole system. It is possible to plot flow-control windows, throughput (packets per unit of time), packet round-trip delays, exponentially weighted average of delay, network feedback (bits, for example), router queue lengths, and router utilization. These time-varying graphs can be observed on line as the simulation is proceeding and are helpful in monitoring the simulation.

For each run, first a set of goal throughputs is computed for each user. This is the throughput that gives the efficient and fair performance for a given configuration and workload. At the end of a batch run, the model computes average (over the run) user throughput and a scaled value (ratio of actual to goal). The variance (across users) of the scaled throughput indicates fairness. For a totally fair system, the variance should be zero. The average (across users) of the scaled throughput indicates efficiency. For an optimal system, the average scaled throughput must be one, i.e., equal to the goal.

Averages and variances of each performance variable is obtained in three different stages. Consider, for example, user throughput. $T(t, u, i, p)$ denotes throughput at time $t$ for $u^{th}$ user in the $i^{th}$ repetition of the simulation with $p^{th}$ input parameter set. $T(u, i, p)$ would represent average throughput for the $u^{th}$ user during $i^{th}$ repetition. $T(i, p)$ would represent average throughput across all users during $i^{th}$ repetition, and finally $T(p)$ would represent average throughput (across all users and replications) for the given parameter set. Variance is similarly computed at each level. High variance over time or high variance across users or high variance over repetition is undesirable. Time variance of flow-control windows gives an idea of the oscillations.

## 10.9 ASSUMPTIONS

### 10.9.1 No Loss of Packets

In our congestion avoidance studies, we concentrated on a *no loss* case. We assumed that the routers have a large buffer capacity. The performance degradation is caused not by packet losses (as was the case with previous congestion control studies) rather by increased queueing in the network.

### 10.9.2 Equal Paths

All results presented in the first two parts of this report series assume *equal paths*, i.e., packets from all sources join the path at first router and leave it at the last router.

## 10.10 LIMITATIONS

The simulation model has the following known limitations.

### 10.10.1 No Reverse Traffic

The acknowledgments traveling back from the destination to the source are not explicitly simulated. The source is informed of the packet delivery as soon as the packet is accepted by the destination.

### 10.10.2 No Acknowledgment Withholding

Each packet passing through the network brings with it the feedback. The acknowledgments returning from the destination bring the feedback back to the source. If there is acknowledgment-withholding so that a single acknowledgment is sent for a number of packets received, the destination would need an algorithm to combine multiple feedbacks received or to piggyback all of them on the single acknowledgment. We have not yet studied such cases.

## 11  TEST SEQUENCE

In order to verify that the congestion avoidance algorithms that we developed do satisfy the goals described earlier in this report, we created a number of test scenarios for simulation. In this section, we describe these test scenarios and show how they help verify the goals. For each alternative considered, we repeated the test sequence, and if the alternative did not provide acceptable performance for a test case, the alternative was rejected.

The test sequences are identified by a three or four character identifier constructed as follows. The first character indicates whether the path is Homogeneous (H), Non-Homogeneous (N), Satellite Link (S), or Mixed (M). A non-homogeneous path with one or more satellite links is called mixed. The second character in the identifier indicates if the packet lengths are Deterministic (D) or Random (R). The third character $n$ is a digit identifying the number of users sharing the path. The fourth character, which is optional, indicates special conditions and will be described later.

### 11.1  EFFICIENCY

Most alternatives were first tested for the MD1 configuration shown in Figure 9. This is a mixed configuration with deterministic packet lengths and a single user. It is clear that the alternatives that do not work with single users are not worth further consideration. The deterministic nature of the simulation helps identify the bugs in the implementation of the alternative. Due to the presence of the satellite delay, the efficient (knee) window is generally sufficiently high to help clearly see the efficiency.

The single-user case also helps identify optimal values of some parameters. The bursty nature of network traffic makes the single-user case performance the most important.

If the alternative works satisfactorily for MD1, we test other sub-cases such as ND1 (without the satellite) and HD1 (homogeneous). We found a few schemes that work mainly because of the long delay introduced by the satellite and therefore do not work with ND1 and HD1.

**Figure 9:** MD1 configuration used for testing efficiency.

**Figure 10:** MDn configuration used for testing fairness.

## 11.2 FAIRNESS

The next step is to see if the alternative is fair. We therefore move to the MDn configuration shown in Figure 10 which is similar to MD1 configuration shown earlier but with $n$ users sharing the same path. In a totally fair scheme, each user should get $(1/n)^{th}$ of the total system throughput.

**Figure 11:** ND9 configuration used for testing overload case.

## 11.3 CONVERGENCE

Some schemes diverge if there are too many users. Such schemes were discovered by using an ND9 configuration shown in Figure 11. This configuration has nine users sharing a non-homogeneous deterministic path. For this configuration, we set the service rates such that the knee occurs when the sum of the window sizes is 3. Obviously, with nine users each having a minimum window limit of 1, the knee can never be achieved. The best alternatives keep the window fixed at 1 for all users. Others allow them to alternate between 1 and 3. Divergent alternatives allowed the window sizes to grow indefinitely in this test case.

## 11.4 ROBUSTNESS

The scheme should work satisfactorily for any given distribution of packet sizes. We found the exponential distribution to be the one most difficult to satisfy. Therefore, we test robustness by using MR1 and NR1 configurations with exponentially distributed packet lengths. Single-user cases help us concentrate on the effects of randomness. We replicated the simulation several times (with different random-number generation seeds) to verify that the scheme does give an acceptable average performance and that the variation of the window is acceptable. NR1 configuration without satellite is more difficult than MR1 because in the latter, the satellite delay, which constitutes a large part of the total delay, is a constant.

## 11.5 RESPONSIVENESS

We test responsiveness by introducing temporary changes in the bottleneck router speed and verifying that the scheme adapts to the new knee. This test case is identified as MD1T, a mixed, deterministic, single-user case with transients. The transient feature of the simulation was described earlier in this report.

Two other aspects of responsiveness are starting credit value and staggered start. These are described separately below.

## 11.6 ANY INITIAL WINDOW

If a scheme is responsive and adapts to changes in the network configuration, the initial window at which a user starts should not matter. We verify this requirement by using an MD1H configuration in which the user starts at a very high window, generally several times the knee value.

## 11.7 STAGGERED START

Another aspect of responsiveness is the change in the number of users. In actual networks, users constantly get on/off the network. We simulate this condition by using the staggered start feature of the simulation. The configuration called MDnS starts initially with one user and the $(i+1)^{st}$ user comes on after $i^{th}$ user has sent a prespecified fraction, for instance 10%, of its total packets.

## 11.8 SOURCE BOUND

It is possible to have networks in which the sources of traffic, not the networks, are the bottleneck. They cannot send packets as fast as the network would allow them to. We call this the source-bound case and it is tested by the configuration called ND1B - non-homogeneous deterministic, one user with bottleneck source.

# 12 SUMMARY

The key contributions of the research reported in this report are the following.

1. We have introduced a new term *congestion avoidance.* It has been distinguished from other similar terms of flow control and congestion control. It is shown that the preventive nature of congestion avoidance helps the network use its resources in a globally optimal manner.

2. We have shown that the key to congestion avoidance in connectionless networks is feedback from the network to the users about its state and a cooperative effort on the part of the users to adjust their demands based on this feedback.

3. We listed a number of alternatives for the feedback and described the criteria used to select the set of acceptable alternatives. Also, a number of goals were described that helped select the best alternative from this set.

4. It has been shown that the problem of congestion avoidance can be broken down into a number of components. The interrelationships among these components, if any, were described.

5. The demand (flow-control window) increase/decrease algorithms used by sources play a key role in the design of a congestion avoidance scheme. These algorithms, as well as the question of demand update frequency, were discussed.

6. The goals that we set to measure the goodness of congestion avoidance schemes were quantified using a simulation model and a series of test sequences. These were described.

This is the first report in a series describing our research on congestion avoidance. The next report in this series [19] describes the application of the concepts described here to the development of a congestion avoidance scheme using binary feedback.

## 13   ACKNOWLEDGMENTS

Many architects and implementers of Digital's networking architecture participated in a series of meetings over the last three years in which the ideas presented here were discussed and improved. Almost all members of the architecture group contributed to the project in one way or another. In particular, we would like to thank Tony Lauck and Linda Wright for encouraging us to work in this area. Radia Perlman, Art Harvey, Kevin Miles, and Mike Shand are the responsible architects whose willingness to incorporate our ideas provided further encouragement. We would also like to thank Bill Hawe, Dave Oran, and John Harper for feedback and interest. The idea of proportional decrease was first proposed by George Verghese. The concept of maximal fairness was proposed by Bob Thomas.

# References

[1] V. Ahuja, "Routing and Flow Control in Systems Network Architecture," IBM Systems Journal, Vol. 18, No. 2, 1979, pp. 298 - 314.

[2] K. Bharat-Kumar and J. M. Jaffe, "A New Approach to Performance-Oriented Flow Control," IEEE Transactions on Communications, Vol. COM-29, No. 4, April 1981, pp. 427 - 435.

[3] W. Bux and D. Grillo, "Flow Control in Local-Area Networks of Interconnected Token Rings," IEEE Transactions on Communications, Vol. COM-33, No. 10, October 1985, pp. 1058-66.

[4] Dah-Ming Chiu and Raj Jain, "Congestion Avoidance in Computer Networks with a Connectionless Network Layer. Part III: Analysis of Increase/Decrease Algorithms," Digital Equipment Corporation, Technical Report #TR-509, August 1987.

[5] David Clark, "NETBLT: A Bulk Data Transfer Protocol," Massachusetts Institute of Technology, Lab for Computer Science, RFC-275, February 1985.

[6] Digital Equipment Corp., "DECnet Digital Network Architecture NSP Functional Specification, Phase IV, Version 4.0.0," March 1982.

[7] M. Gerla and L. Kleinrock, "Flow Control: A Comparative Survey," IEEE Transactions on Communications, Vol. COM-28, No. 4, April 1980, pp. 553 - 574.

[8] A. Giessler, J. Haanle, A. Konig and E. Pade, "Free Buffer Allocation - An Investigation by Simulation," Computer Networks, Vol. 1, No. 3, July 1978, pp. 191-204.

[9] International Organization of Standardization, "ISO 8073: Information Processing Systems - Open Systems Interconnection - Connection Oriented Transport Protocol Specification," July 1986.

[10] J. M. Jaffe, "Flow Control Power is Nondecentralizable," IEEE Transaction on Communications, Vol. COM-29, No. 9, September 1981, pp. 1301-1306.

[11] Raj Jain, Dah-Ming Chiu, and William Hawe, "A Quantitative Measure of Fairness and Discrimination for Resource Allocation in Shared Systems," Digital Equipment Corporation, Technical Report TR-301, September 1984.

[12] Raj Jain, "Using Simulation to Design a Computer Network Congestion Control Protocol," Proc. Sixteenth Annual Modeling and Simulation Conference, Pittsburgh, PA, April 1985.

[13] Raj Jain, "Divergence of Timeout Algorithms for Packet Retransmission," Proc. Fifth Annual International Phoenix Conf. on Computers and Communications, Scottsdale, AZ, March 26-28, 1986, pp. 174-179.

[14] Raj Jain, "A Timeout-Based Congestion Control Scheme for Window Flow-Controlled Networks," IEEE Journal on Selected Areas in Communications, Vol. SAC-4, No. 7, October 1986, pp. 1162-1167.

[15] J. M. McQuillan, I. Richer, and E. C. Rosen, "The New Routing Algorithm for the ARPANET," IEEE Transactions on Communications, Vol. COM-28, No. 5, May 1980, pp. 711-719.

[16] J.C. Majithia, et al., "Experiments in Congestion Control Techniques," Proc. Int. Symp. Flow Control Computer Networks, Versailles, France. February 1979.

[17] John Nagle, "Congestion Control in TCP/IP Internetworks," Computer Communication Review, Vol. 14, No. 4, October 1984, pp. 11-17.

[18] K. K. Ramakrishnan, "Analysis of a Dynamic Window Congestion Control Protocol in Heterogeneous Environments Including Satellite Links," Proceedings of Computer Networking Symposium, November 1986.

[19] K. K. Ramakrishnan and Raj Jain, "Congestion Avoidance in Computer Networks with a Connectionless Network Layer. Part II: An Explicit Binary Feedback Scheme," Digital Equipment Corporation, Technical Report #TR-508, August 1987.

[20] A.S. Tanenbaum, Network Protocols. Prentice-Hall: Englewood Cliffs, NJ, 1981.